\documentclass[aps,preprint,epsfig,rotate]{revtex4}
\usepackage{graphicx}
\usepackage{bm}
\usepackage{epsfig}

\input epsf
\begin{document}
\title{On contribution of the electron-electron correlations into bremsstrahlung from few-electron ions}

 \author{Alexei M. Frolov}
 \email[E--mail address: ]{afrolov@uwo.ca}

\affiliation{Department of Applied Mathematics \\
 University of Western Ontario, London, Ontario N6H 5B7, Canada}

\author{David M. Wardlaw}
 \email[E--mail address: ]{dwardlaw2@mun.ca}

\affiliation{Department of Chemistry, Memorial University of Newfoundland, St.John's, 
             Newfoundland and Labrador, A1C 5S7, Canada}

\date{\today}

\begin{abstract}

A new approach is developed to evaluate contribution of the electron-electron correlations into bremsstrahlung from few-electron ions and atoms. Our approach is based on 
the explicit formula for the electron density distribution in such systems. We derive the closed analytical formula for the matrix elements which are needed for highly 
accurate computations of atomic form-factors of two-electron atoms and ions. We also discuss the energy loss due to bremsstrahlung in a plasma which contains multi-charged 
ions and free electrons. Bremsstrahlung from a high-temperature plasma is considered as well as its role in the high-temperature burn-up of deuterium plasma. 

\end{abstract}

\maketitle
\newpage

\section{Introduction}

As is well known electron-electron correlations in atoms and ions contribute to many bound state properties of these systems. Recently, it appears that such correlations also affect a number 
of properties of atoms and ions, including the influence of contributions from unbound atomic states. In particular, in this study we consider the braking radiation (or bremsstrahlung) 
emitted by a fast electron when it is slowed down, or even stoped, by a few electron ion/atom. In general, this process is represented by the equation \cite{Heitl}: $A^{q+} + e^{-} =  A^{q+} 
+ e^{-} + \hbar \omega$, where the notation $A^{q+}$ stands for a multi-charged (atomic) ion which has a positive electric charge $q+$, while the notations $e^{-}$ and $\hbar \omega$ designate 
the fast electron and emitted photon, respectively. It appears that the bremsstrahlung cross-section and corresponding energy loss explicitly and substantially depend upon the electron-electron
correlations in the incident ions/atom. Our goal in this short study is to investigate this effect and its role for accurate evaluations of the cross-section and energy loss due to bremsstrahlung
in few-electron ions/atoms. Below, we apply the relativistic units in which $\hbar = 1, c = 1$ and $m_e = \alpha^{-1}$, where $\alpha = \frac{e^2}{\hbar c} \approx$ 7.2973525664$\cdot 10^{-3}$ 
\cite{CRC} is the dimensionless fine-sructure constant. The unit of length in these units coincides with the Compton wavelength of the electron $\Lambda_e$, where $\Lambda_e = \frac{\hbar}{m_e c} 
= \alpha a_0 \approx 3.8615926764 \cdot 10^{-11}$ $cm$ and $a_0 = 5.2917721067\cdot 10^{-9}$ $cm$ \cite{CRC} is the Bohr radius.  

In this study the braking radiation emitted during a Coulomb collision of a fast electron with the positively charged atomic nuclues is designated by the German word `bremsstrahlung' \cite{Heitl}. 
Bremsstrahlung has been analyzed with the methods from Classical Electrodynamics and later from Quantum Electrodynamics (see, e.g., \cite{Grein}, \cite{AB} and references therein). The goal of the 
QED analysis was to derive the closed analytical formula(s) for the bremsstrahlung cross-section as a function of the photon frequency $\omega$ and electron's `quantum numbers'. For a `free' 
electron a set of `good (or `convenient') quantum numbers includes the kinetic energy $E_{k}$ and the vector of momemtum ${\bf p}_{k}$, where $k = i$ for the incident electron and $k = f$ for the 
final electron. In actual experiments to measure bremsstrahlung intensity it is very hard to measure angular correlations between directions of propagation of the incident/final electron and emitted 
photon. Therefore, for quantum numbers it is better to use the kinetic energy of the electron $E_{k}$ and absolute value of its momemtum $p_k = \mid {\bf p}_{k} \mid$ (scalar). Below, we shall 
assume that we have an electron with the incident quantum numbers $(E_1, {\bf p}_1)$ which emits one photon $\hbar \omega = \omega$. The final quantum numbers of the electron are $(E_2, {\bf p}_2)$. 
In this notation the well known formula for the bremsstrahlung cross-section obtained from direct analytical QED calculations performed in the Born approximation takes the form (see, e.g., \cite{Heitl}, 
\cite{Grein}, \cite{AB})
\begin{eqnarray}
 \frac{d\sigma}{d \omega} &=& \alpha^{3}  \Lambda^{2}_{e} \Bigl(\frac{Q^2}{\omega}\Bigr) \Bigl[ 1 - \frac{F(q)}{Q} \Bigr]^2 \cdot \frac{p_2}{p_1} \Bigl\{ \frac43 - 2 E_1 E_2 
 \frac{p^2_1 + p^2_2}{p^2_1 p^2_2} + m^2 \Bigl[ \frac{E_2}{p^3_1} \ln\Bigl(\frac{E_1 + p_1}{E_1 - p_1}\Bigr) \nonumber \\
 &+& \frac{E_1}{p^3_2} \ln\Bigl(\frac{E_2 + p_2}{E_2 - p_2}\Bigr) - \frac{1}{p_1 p_2} \ln\Bigl(\frac{E_1 + p_1}{E_1 - p_1}\Bigr) \ln\Bigl(\frac{E_2 + p_2}{E_2 - p_2}\Bigr)\Bigr]
 + \ln\Bigl(\frac{E_1 E_2 + p_1 p_2 - m^2}{m \omega}\Bigr) \label{generl} \\
 &\times& \Bigl[ \frac83 \frac{E_1 E_2}{p_1 p_2} + \frac{\omega^2}{p^3_1 p^3_2} (E^2_1 E^2_2 + p^2_1 p^2_2) + \frac{m^2 \omega}{2 p_1 p_2} \Bigl( \frac{E_1 E_2 + p^2_1}{p^3_1} 
   \Bigr) \ln\Bigl(\frac{E_1 + p_1}{E_1 - p_1}\Bigr) \nonumber \\
 &+& \frac{m^2 \omega}{2 p_1 p_2} \Bigl( \frac{E_1 E_2 + p^2_2}{p^3_2} \Bigr) \ln\Bigl(\frac{E_2 + p_2}{E_2 - p_2}\Bigr) + \frac{2 \omega E_1 E_2}{p^2_1 p^2_2}\Bigr) 
 \Bigr] \Bigr\} \nonumber 
\end{eqnarray}
where $m = m_e$ is the electron mass, $\Lambda_e$ is the electron's Compton wave length ($a_0 = \alpha^{-1} \Lambda_{e}$ is the Bohr radius) and $Q$ is the electric charge of the atomic nucleus in an 
ion/atom which has $N_e$ bound electrons, where $N_e \le Q$. The factor $F(q)$ used in this formula is the so-called form-factor of the ion/atom. This atomic form-factor is defined traditionally as:
\begin{eqnarray}
   F(q) = \int_{0}^{+\infty} \oint n_e(r) \exp(-{\bf q} \cdot {\bf r}) r^2 dr d\Omega = 4 \pi \int_{0}^{+\infty} n_e(r) \sin(q r) r dr \label{formf}
\end{eqnarray}
where $n_e(r)$ is the electron density in a few-electron atom/ion which is assumed to be spherically symmetric. The density $n_e(r)$ is normalized and its norm equals the number of bound electrons 
$N_e$, i.e. $\int_{0}^{+\infty} n_e(r) r^2 dr = N_e$. The parameter $q$ in the expression for the form-factor $F(q)$ equals to the ratio of the velocity of the fast electron $v_e$ to the atomic 
velocity $\alpha c (= 1$ in the relativistic units). 

At small non-relativistic energies, when $p_1 \ll m c = m$ in relativistic units, the formula, Eq.(\ref{generl}), takes the form
\begin{eqnarray}
  \frac{d\sigma}{d \omega} = \alpha^{3} \Lambda^{2}_{e} \Bigl(\frac{Q^2}{\omega}\Bigr) \Bigl[ 1 - \frac{F(q)}{Q} \Bigr]^2 \cdot \frac{m}{T_1} \ln\Bigl[\frac{(\sqrt{T_1} + 
   \sqrt{T_1 - \omega})^2}{\omega}\Bigr]  \label{crsnonrel}
\end{eqnarray}
where $T_1 = \frac{p^2_1}{2 m}$ is the non-relativistic kinetic energy of the incident electron. Note that this expression does not contain any energy and/or momentum of the
final electron, i.e any of the $E_2$ and/or $p_2$ values. In the opposite ultra-relativistic case, when $E_1 \gg m$ and $E_2 \gg m$, the formula for the bremsstrahlung 
cross-section, Eq.(\ref{generl}), is written in the form
\begin{eqnarray}
  \frac{d\sigma}{d \omega} = 4 \alpha^5 a^2_0 \Bigl(\frac{Q^2}{\omega}\Bigr) \Bigl[ 1 - \frac{F(q)}{Q} \Bigr]^2 \cdot \frac{E_2}{E_1} \Bigl( \frac{E_1}{E_2} + 
  \frac{E_2}{E_1} - \frac23 \Bigr) \cdot \Bigl[ \ln\Bigl(\frac{2 E_1 E_2}{m \omega}\Bigr) - \frac12 \Bigr]  \label{crsrel}
\end{eqnarray}
where $\alpha^{3} \Lambda^{2}_{e} = \alpha^5 a^2_0$ and $E_2 = E_1 - \omega$ and $\alpha^5 a^2_0 \approx 5.79467274 \cdot 10^{-28}$ $cm^{2}$. It is interesting to note that the factor 
$\alpha^5 a^2_0 \Bigl(\frac{Q^2}{\omega}\Bigr) \Bigl[ 1 - \frac{F(q)}{Q} \Bigr]^2$ in Eq.(\ref{generl}) and Eqs.(\ref{crsnonrel}) - (\ref{crsrel}) can be written in the following form
\begin{eqnarray}
 4 \alpha^5 a^2_0 \Bigl(\frac{Q^2}{\omega}\Bigr) \Bigl[ 1 - \frac{F(q)}{Q} \Bigr]^2 = \alpha^5 [a_0 q^2 f_B(q)]^2
\end{eqnarray}
where $f_B(q)$ is the Born amplitude for elastic scattering of an electron by a $N_e$-electron ion/atom (see, e.g., \cite{LLQ} and \cite{Bethe}). In the case of elastic scattering $q = 
2 k \sin\frac{\theta}{2}$. This means that by determining the scattering amplitude of elastic scattering in the Born approximation one also finds the factor which is needed for 
calculation of the differential cross-section of bremsstrahlung $\frac{d\sigma}{d \omega}$ in the ultra-relativistic limit. 

Note that each of the formulas Eq.(\ref{generl}) and Eqs.(\ref{crsnonrel}) - (\ref{crsrel}) contains the form-factor $F(q)$ defined in Eq.(\ref{formf}). For many few-electron ions the 
overall contribution of the form-factor is relatively small and rapidly decreases when the nuclear charge $Q$ grows. Moreover, for such atomic systems the electron density 
$n_e({\bf r})$ can be considered (to very high accuracy) as a spherically symmetric function. This corresponds to the actual physical picture of bremsstrahlung as a braking radiation 
emitted during Coulomb interaction between a fast electron and atomic nucleus. The role of atomic electrons is an electrical screening of the central nucleus. In general, atomic 
form-factors for different atoms/ions can be determined to high accuracy by using computational methods of modern atomic physics. Derivation of the analytical formulas and numerical 
calculations of the form-factors for several different few-electron ions and atoms is one of the goals of this study. 

\section{Formulas for the form-factor in few-electron ions}

As follows from the formulas given in the Introduction to determine the bremsstrahlung cross-sections for actual few-electron ions/atoms one needs to evaluate the atomic form-factor 
$F(q)$, Eq.(\ref{formf}). In this Section we derive closed analytical expressions for this quantity in different few-electron ions/atoms. For one-electron atomic systems we can use 
hydrogenic wave functions. This drastically simplifies all calculations of the atomic form-factor for different bound states, i.e. states with the different values of the angular momentum 
$L$ and the principal quantum number $n$. To simplify the problem let us consider the ground, doublet $1^2S-$state in hydrogen-like, one-electron ions with nuclear charge $Q$. In this case 
the form-factor $F(q)$ is
\begin{eqnarray}
 F(q) &=& 4 Q^3 \int_{0}^{+\infty} \exp(-2 Q r) \sin(q r) r dr = \frac{4 Q^3 \Gamma(2)}{4 Q^2 + q^2} \sin\Bigl[2 \arctan\Bigl(\frac{q}{2 Q}\Bigr)\Bigr] \label{ffhydr} \\
   &=& \frac{8 Q^3}{4 Q^2 + q^2} \sin\Bigl[ \arcsin \frac{\frac{q}{2 Q}}{\sqrt{1 + \Bigl(\frac{q}{2 Q}\Bigr)^2}} \Bigr] 
   \cos\Bigl[ \arccos \frac{1}{\sqrt{1 + \Bigl(\frac{q}{2 Q}\Bigr)^2}} \Bigr] = \frac{16 q Q^4}{(4 Q^2 + q^2)^2} \nonumber
\end{eqnarray}
where we used the formula Eq.(3.944.5) from \cite{GR}. This formula can easily be generalized to the case of two-electron ions/atoms, if it is possible to neglect the electron-electron 
correlations in such systems. In the lowest order approximation such a generalization can be achieved with the substitution $Q \rightarrow Q - \frac{5}{16}$ in the formula, Eq.(\ref{ffhydr}). 
The atomic form factor changes correspondingly. However, the first term in the $[Q - F(q)]^2$ expression, i.e. the electric charge of the nucleus $Q$, which can be found in each of 
Eq.(\ref{generl}) and Eqs.(\ref{crsnonrel}) - (\ref{crsrel}) does not change, since $Q$ is the actual nuclear charge in the ions. These arguments lead to the following approximate formula for 
two-electron multi-charged ions
\begin{eqnarray}
 [ Q - F(q) ]^2 = Q^2 \Bigl\{1 - \frac{16 N_e q (Q - \frac{5}{16})^4}{Q^2 [4 (Q - \frac{5}{16})^2 + q^2]^2} \Bigr\}^2 =  Q^2 \Bigl\{1 - \frac{32 q (Q - \frac{5}{16})^4}{Q^2 [4 (Q - \frac{5}{16})^2 
 + q^2]^2} \Bigr\}^2 \label{ffhelike}
\end{eqnarray}
where $N_e = 2$ is the total number of bound electrons. These calculations are simple and straighforward. However, analogous computations with the use of the truly correlated wave functions
for few-electron ions/atoms become significantly more complicated.

For relative simplicity, let us consider the truly correlated wave functions of two-electron atoms/ions. Such wave functions can be represented in terms of the different variational expansions. 
One of the most effective and accurate expansion is the exponential variational expansion written in relative/perimetric coordinates. For the ground $1^1S-$states this expansion takes the form
(see, e.g., \cite{Fro98} and \cite{Fro2001})
\begin{eqnarray}
 \Psi &=& \Bigl( 1 + \hat{P}_{12} \Bigr) \sum_{i=1}^{N} C_{i} \exp(-\alpha_{i} r_{32} - \beta_{i} r_{31} - \gamma_{i} r_{21}) \label{exp1} \\
 &=& \sum_{i=1}^{N} C_{i} \Bigl[ \exp(-\alpha_{i} r_{32} - \beta_{i} r_{31} - \gamma_{i} r_{21}) + \exp(-\beta_{i} r_{32} - \alpha_{i} r_{31} - \gamma_{i} r_{21})
  \Bigr] \nonumber 
\end{eqnarray}
The expansion, Eq.(\ref{exp1}), is called the exponential variational expansion in the relative coordinates $r_{32}, r_{31}$ and $r_{21}$. Each of the three relative coordinates $r_{ij}$ is defined 
as the difference between the corresponding Cartesian coordinates of the two particles, i.e., $r_{ij} = \mid {\bf r}_i - {\bf r}_j \mid = r_{ji}$. It follows from this definition that the relative 
coordinates $r_{32}, r_{31}$ and $r_{21}$ are translationally and rotationally invariant. Below, the index 3 is used to designate the atomic nucleus, while indexes 1 and 2 stand for atomic electrons.
The coefficients $C_i$ are the linear variational parameters of the expansion, Eq.(\ref{exp1}), while the parameters $\alpha_{i}, \beta_{i}$ and $\gamma_{i}$ are the non-linear parameters of this 
expansion. In general, the total energy of the ground $1^1S-$state of the two-electron ion depends upon the total number of basis functions $N$ used in calculations. The operator $\hat{P}_{12}$ in 
Eq.(\ref{exp1}) is the permutation operator for two identical particles (electrons). The very high efficiency of the variational expansion, Eq.(\ref{exp1}), in actual applications to the two-electron 
ions is related to the fact that all non-linear parameters $\alpha_{i}, \beta_{i}$ and $\gamma_{i}$ are carefully varied in such calculations. 

An analytical formula for the form-factor $F(q)$ derived with the use of the exponential variational expansion takes the following form
\begin{eqnarray}
   F(q) &=& \frac12 \sum_{i=1}^{N} \sum_{j=1}^{N} C_{i} C_{j} \Bigl[ \int_{0}^{\infty} r_{32} dr_{32} \sin(q r_{32}) \exp[-(\alpha_i + \alpha_j) r_{32}] 
    \int_{0}^{\infty} r_{31} dr_{31} \exp[-(\beta_i + \beta_j) r_{31}] \times \nonumber \\
  & & \int_{|r_{32} - r_{31}|}^{r_{32} + r_{31}} r_{21} dr_{21} \exp[-(\gamma_i + \gamma_j) r_{21}] + \int_{0}^{\infty} r_{32} dr_{32} \sin(q r_{32}) \exp[-(\alpha_i + \beta_j) r_{32}]
   \label{exp2} \\
 & &  \int_{0}^{\infty} r_{31} dr_{31} \exp[-(\beta_i + \alpha_j) r_{31}] \int_{|r_{32} - r_{31}|}^{r_{32} + r_{31}} r_{21} dr_{21} \exp[-(\gamma_i + \gamma_j) r_{21}] \nonumber
\end{eqnarray}
First, we can calculate the internal integral in this equation. The result is written in the two following forms: (1) for $r_{32} \ge r_{31}$
\begin{eqnarray}
  I&=& \int_{|r_{32} - r_{31}|}^{r_{32} + r_{31}} r_{21} dr_{21} \exp[-(\gamma_i + \gamma_j) r_{21}] = 
    \exp[-(\gamma_i + \gamma_j)(r_{32} - r_{31})] \Bigl(\frac{r_{32} - r_{31}}{\gamma_{i} + \gamma_{j}} + \frac{1}{(\gamma_{i} + \gamma_{j})^2}\Bigr) \nonumber \\ 
  &-& \exp[-(\gamma_i + \gamma_j)(r_{31} + r_{32})] \Bigl(\frac{r_{32} + r_{31}}{\gamma_{i} + \gamma_{j}} + \frac{1}{(\gamma_{i} + \gamma_{j})^2}\Bigr) \label{exp3a}
\end{eqnarray}
and (2) for $r_{32} \le r_{31}$
\begin{eqnarray}
 J&=& \int_{|r_{32} - r_{31}|}^{r_{32} + r_{31}} r_{21} dr_{21} \exp[-(\gamma_i + \gamma_j) r_{21}] = 
  \exp[-(\gamma_i + \gamma_j)(r_{31} - r_{32})] \Bigl(\frac{r_{31} - r_{32}}{\gamma_{i} + \gamma_{j}} + \frac{1}{(\gamma_{i} + \gamma_{j})^2}\Bigr) \nonumber \\
 &-& \exp[-(\gamma_i + \gamma_j)(r_{31} + r_{32})] \Bigl(\frac{r_{31} + r_{32}}{\gamma_{i} + \gamma_{j}} + \frac{1}{(\gamma_{i} + \gamma_{j})^2}\Bigr) \label{exp3b}
\end{eqnarray}
These two integrals $I$ and $J$ are the functions of the two variables $r_{32} + r_{31}$ and $|r_{32} - r_{31}|$. By performing integration over the $r_{31}$ relative coordinate one obtains the explicit 
formula for the form-factor $F(q)$ as a quantity which depends upon the $r = r_{32}$ radial variable (or electron-nucleus distance $r_{eN}$). In some cases the knowledge of this dependence is crucial
to understand the nature of the physical problem. 

However, if we need to know only the absolute value of the form-factor $F(q)$, then it is possible to use another approach to its calculation. In this case instead of three relative interparticle 
coordinates $r_{32}, r_{31}$ and $r_{21}$ we can introduce three perimetric coordinates $u_1, u_2, u_3$, where $u_{k} = \frac12 (r_{ki} + r_{kj} - r_{ij})$, $r_{ab} = r_{ba}$ and $(i, j, k)$ = 
(1, 2, 3) \cite{Fro98}. The inverse relation take the form $r_{ab} = u_a + u_b$. The three perimetric coordinates generally are independent of each other, non-negative and each of them varies between 
0 and $+\infty$. As follows from these properties, three perimetric coordinates form a very convenient set of variables in order to calculate arbitrary three-particle integrals. Let us apply perimetric 
coordinates  to determine the atomic form-factors of the different two-electron ions/atoms. First, consider the integral between $j_{0}(q r_{32})$ and $i$ and $j$ exponential basis functions. In 
relative coordinates we can write
\begin{eqnarray}
 I &=& \int_{0}^{\infty} \int_{0}^{\infty} \int_{|r_{32} - r_{31}|}^{r_{32} + r_{31}} \frac{\sin(q r_{32})}{q r_{32}} \exp[-(\alpha_i + \alpha_j) r_{32} -(\beta_i + \beta_j) r_{31} -(\gamma_i + 
 \gamma_j) r_{21}] r_{32} r_{31} \times \nonumber \\
 & & r_{21} dr_{32} dr_{31} dr_{21} \label{integ1}
\end{eqnarray}  
where the factor $j_{0}(q r_{r_2}) = \frac{\sin(q r_{32})}{q r_{32}}$ is the spherical Bessel function of zero order. In perimetric coordinates this integral takes the form 
\begin{eqnarray}
 I &=& \frac{2}{q} \int_{0}^{\infty} \int_{0}^{\infty} \int_{0}^{\infty} \sin[q (u_3 + u_2)] \exp[- Z u_3 - Y u_2 - X u_1] (u_3 + u_1) (u_2 + u_1) du_3 du_2 du_1 \nonumber \\
 &=& \frac{2}{q} \int_{0}^{\infty} \int_{0}^{\infty} \int_{0}^{\infty} [\sin(q u_3) \cos(q u_2) + \cos(q u_3) \sin(q u_2)] (u_3 u_2 + u_3 u_1 + u_2 u_1 + u^2_1) \times \nonumber \\ 
 & & \exp[- Z u_3 - Y u_2 - X u_1] du_3 du_2 du_1 \label{integ2} 
\end{eqnarray} 
where $Z = \alpha_i + \alpha_j + \beta_i + \beta_j, Y = \alpha_i + \alpha_j + \gamma_i + \gamma_j$ and $X = \beta_i + \beta_j + \gamma_i + \gamma_j$. The factor 2 in this formula is the Jacobian of the 
transformation $(r_{32}, r_{31}, r_{21}) \rightarrow (u_3, u_2, u_1)$. As follows from Eq.(\ref{integ2}), calculation of the integral $I$ is reduced to the analytical computation of eight integrals, where 
each of the contributing integrals is the product of three one-dimensional integrals. For instance, the first contributing integral is
\begin{eqnarray}
 I_1 &=& \frac{2}{q} \int_{0}^{\infty} \int_{0}^{\infty} \int_{0}^{\infty} \sin(q u_3) \cos(q u_2) \exp[- Z u_3 - Y u_2 - X u_1] u_3 u_2 du_3 du_2 du_1 \label{integ31} \\
 &=& \frac{2}{q X} \int_{0}^{\infty} \sin(q u_3) \exp[- Z u_3] u_3 du_3 \cdot \int_{0}^{\infty} \cos(q u_2) \exp[- Y u_2] u_2 du_2 = \frac{2 Z (Y^2 - q^2)}{X (Z^2 + q^2) 
  (Y^2 + q^2)^2} \nonumber
\end{eqnarray} 
while the last integral $I_8$ is  
\begin{eqnarray}
 I_8 &=& \frac{2}{q} \int_{0}^{\infty} \int_{0}^{\infty} \int_{0}^{\infty} \cos(q u_3) \sin(q u_2) \exp[- Z u_3 - Y u_2 - X u_1] u^2_1 du_3 du_2 du_1 \label{integ38} \\
 &=& \frac{4}{q X^3} \int_{0}^{\infty} \cos(q u_3) \exp[- Z u_3] du_3 \cdot \int_{0}^{\infty} \sin(q u_2) \exp[- Y u_2] du_2 = \frac{4 Z}{X^3 (Z^2 + q^2) (Y^2 + q^2)} \nonumber
\end{eqnarray} 
The final formula for the $(ij)$-matrix element of the form-factor $F(q)$ is 
\begin{eqnarray}
  [ F(q) ]_{ij} &=& \frac{2}{X (Z^2 + q^2) (Y^2 + q^2)} \Bigl[ \frac{2 Z (Y^2 - q^2)}{(Z^2 + q^2) (Y^2 + q^2)} + \frac{2 Y (Z^2 - q^2)}{(Z^2 + q^2) (Y^2 + q^2)} + \frac{2 Y Z}{X (Z^2 + q^2)} \nonumber \\
  &+& \frac{Z^2 - q^2}{X (Z^2 + q^2)} + \frac{Y^2 - q^2}{X (Y^2 + q^2)} + \frac{2 Y Z}{X (Y^2 + q^2)} + \frac{2 Y}{X^2} + \frac{2 Z}{X^2}\Bigr] \label{fform}
\end{eqnarray} 

It is clear that this formula is a regular function of $q$ (i.e., it contains no singularities) and is numerically stable for arbitrary $q$. The formula, Eq.(\ref{fform}), for matrix elements is similar to 
our formulas derived in \cite{Fro2015} for the matrix elements involving spherical Bessel functions which have been obtained with the use of the same approach. Results of highly accurate numerical 
calculations of some atomic form-factors can be found in Tables I and II (for different ions/atoms and different values of $q$). Table I shows convergece of the form-factor for the ground $1^1S-$state of 
the ${}^{\infty}$He-atom determined with the use of the formula, Eq.(16), and different number(s) of basis functions $N$ in Eq.(\ref{exp1}). As follows from Table I convergence of the form-factor $F(q)$ 
upon $N$ in Eq.(\ref{exp1}) is very fast for the He atom (ground $1^1S-$state). Briefly, we can say that results from Table I are almost $N-$independent. The same situation can be obesrved with the 
form-factors of other two-electron ions showh in Table II (H$^{-}$, Ne$^{8+}$, Ca$^{18+}$, Ni$^{26+}$). This indicates the main advantage of our formula, Eq.(\ref{fform}), used for numerical calculations of 
the form-factors $F(q)$ in the two-electron atoms/ions. Formally, it is possible to say that our method based on the formula, Eq.(\ref{fform}), completely solves the problem of highly accurate computations 
of the atomic form-factors for two-electron ions/atoms. 

Table II contains form-factors $F(q)$ (in $a.u.$) determined for a number of two-electron ions/atoms in their ground $1^1S-$states. It is interesting to observe changes of the form-factors for such atomic 
systems due to changes in the electric charge of the nucleus $Q$. For very compact ions, e.g, for the Ca$^{18+}$, Ni$^{26+}$ ions, form-factors change very little when the parameter $q$ varies between 0 and 
11 (see Table III). On the other hand, for the weakly-bound H$^{-}$ ion all substantial changes of the form-factor are located in this area of $q$ variations. The same is true for the form-factor of the 
neutral He atom.    

The third approach for determining the form-factor $F(q)$ is based on the following approximate formula known from atomic physics (see, e.g., \cite{MotMessi}):
\begin{eqnarray}
  F(q) = \int \exp(-\imath {\bf q} \cdot  {\bf r}) n_e({\bf r}) d{\bf r} = \frac{4 \pi}{q} \int_{0}^{+\infty} \frac{\sin(q r)}{r} n_{e}(r) r^2 dr \label{ffq}
\end{eqnarray}
where we have assumed that the electron density distribution is spherically symmetric. At large values of $q$ the integrand in the right-hand side of Eq.(\ref{ffq}) is a fast oscillating function. Therefore, 
the range of large values of the $q r$ variable does not contribute to the form-factor $F(q)$. For small $q r$ values ($qr \le q a_0$, where $a_0$ is the Bohr radius) we can write 
\begin{eqnarray}
  F(q) \approx \frac{4 \pi}{q} \int_{0}^{+\infty} \Bigl( 1 - \frac{(q r)^2}{3!} + \frac{(q r)^4}{5!} - \frac{(q r)^6}{7!} + \frac{(q r)^8}{9!} - \frac{(q r)^{10}}{11!} + \ldots \Bigr) n_{e}(r) r^2 dr 
 \label{ffq1}
\end{eqnarray}
Assuming that the electron density $n_e(r)$ is normalized to the number of bound electrons $N_e$ we can write
\begin{eqnarray}
  F(q) = \frac{N_e}{q} \Bigl( 1 - \frac{q^{2} \langle r^{2} \rangle}{3!} + \frac{q^{4} \langle r^{4} \rangle}{5!} - \frac{q^{6} \langle r^{6} \rangle}{7!} + \frac{q^{8} \langle r^{8} \rangle}{9!} 
  - \frac{q^{10} \langle r^{10} \rangle}{11!} + \ldots \Bigr) \label{ffq2}
\end{eqnarray}
where $r = r_{eN}$ is the electron-nucleus distance (scalar coordinate). In the last formula we also eliminated the factor $4 \pi$ which is compensated by the corresponding factors from angular parts of the 
wave functions, or electron density distribution. This means that numerical computations of form-factors for few-electrons atoms and ions are reduced to accurate calculations of the $\langle r^{2k}_{eN} 
\rangle$ expectation values for $k = 1, 2, 3, \ldots$. In reality, for two-electron ions/atoms it is possible to evaluate atomic form-factors by using a finite number of terms, e.g., three, or four terms, in 
Eq.(\ref{ffq2}), since in this case for relatively small $q (\le 1)$ the series, Eq.(\ref{ffq2}), converges very fast. Some of the $\langle r^{2k}_{eN} \rangle$ expectation values determined for a few 
two-electron ions/atoms (in atomic units)  can be found in Table III. Accurate computations of the atomic form-factors for three- and four-electron atoms/ions can be perfomed analogously. Details of such 
calculations and analogous computations of the form-factors for three- and four-electron ions/atoms will be discussed in future studies.

\section{Energy loss due to bremsstrahlung}

The improtance of the derivatives $\frac{d\sigma}{d \omega}$ of the bremsstrahlung cross-section defined by Eq.(\ref{generl}) and Eqs.(\ref{crsnonrel}) - (\ref{crsrel}) follows from the fact 
that these values are directly related to the radiative energy loss, or, in other words, to the energy loss due to bremsstrahlung. Indeed, let us assume that a fast electron with the incident 
quantum numbers $(E_1, {\bf p}_1)$ moves through matter which contains $N_i$ ions/atoms per $cm^3$. Such an electron emits a photon $\hbar \omega = \omega$ and becomes the final electron with 
quantum numbers $(E_2, {\bf p}_2)$. The average energy loss due to bremsstrahlung per $cm$ of electron path $x$ is given by  
\begin{eqnarray}
 -\Bigl(\frac{d E_1}{dx}\Bigr)_{r} = N_i \cdot \int_{0}^{E_1 - m} \omega \Bigl(\frac{d \sigma}{d \omega}\Bigr) d\omega = N_i E_1 \Bigl[ \frac{1}{E_1} \cdot \int_{0}^{E_1 - m} \omega 
  \Bigl(\frac{d \sigma}{d \omega}\Bigr) d\omega \Bigr] = N_i E_1 \sigma_r \; \; \; , \; \; \; \label{loss1}
\end{eqnarray}
where the value $\frac{d \sigma}{d \omega}$ is the differential cross-section of bremsstrahlung defined in the Introduction. The radiative cross-section $\sigma_r$ defines the total radiative 
losses related to bremsstrahlung. It is interesting to note that this formula can be applied to a large number of actual systems, e.g., to evaluate radiative loss in hot plasmas and evaluate 
an additional heating in nuclear reactors which contain substantial amounts of fast decaying $\beta^{-}$ isotopes. In fact, our original interest in this problem was related to numerical 
evaluation of the energy loss due to bremsstrahlung in a sample containing a large number of $\beta^{-}$ decaying nuclei. In an actual fuel rod which has been taken from a working nuclear 
reactor 1 $cm^3$ usually contains between $\approx 1 \cdot 10^{8}$ and $5 \cdot 10^{14}$ of $\beta^{-}$ decaying nuclei. The exact number depends upon the age of the reactor, its type, working 
hystory and other factors. The total number of $\beta^{+}$ decaying nuclei is $\approx$ 15 - 20 smaller than the analogous number of $\beta^{-}$ 
nuclei. 

By using the analytical expression for the $\frac{d \sigma}{d \omega}$ derivative, Eq.(\ref{generl}), we can determine the radiative cross-section $\sigma_r$ in Eq.(\ref{loss1}). The final
formula is 
\begin{eqnarray}
 \sigma_r &=& \alpha^5 a^2_0 \Bigl(\frac{Q^2}{\omega}\Bigr) \Bigl[ 1 - \frac{F(q)}{Q} \Bigr]^2 \Bigl\{ \frac{12 E^2_1 + 4 m^2}{3 E_1 p_1} \ln\Bigl(\frac{E_1 + p_1}{m}\Bigr) 
  - \frac{8 E_1 + 6 p_1}{3 E_1 p^2_1} \Bigl[\ln\Bigl(\frac{E_1 + p_1}{m}\Bigr)\Bigr]^2 - \frac43 \nonumber \\
 &+& \frac{2 m^2}{E_1 p_1} L\Bigl(\frac{2 E_1 (E_1 + p_2)}{m^2}\Bigr)\Bigr\} \; \; \; , \; \; \; \label{loss2}
\end{eqnarray}
where the function $L(x) = -dilog(x+1) = \int_{0}^{x} \Bigl(\frac{1 + y}{y}\Bigr) dy = \frac{\pi^2}{6} + \int_{0}^{x-1} \Bigl(\frac{1 + y}{y}\Bigr) dy = \frac{\pi^2}{6} - dilog(x) = Li_2(-x)$ is 
proportional to the well known dilogarithm function defined by Eq.(27.7.1) in \cite{Abram}. Note that definitions of the dilogarithm function in different books and text books differ from each other. 
In particular, our function $L(x)$ from Eq.(\ref{loss2}) coincides almost exactly with the dilogaritm $Li_2(x)$ defined in $Wolfram$ $Math$ $World$ \cite{Wolfr}, i.e. $L(x) = Li_2(-x)$. Furthermore, 
for our $L(x) = Li_2(-x)$ function one finds:
\begin{eqnarray}
  L(x) = \frac{\pi^2}{6} + \frac{(\ln x)^2}{2} - F\Bigl(\frac{1}{x}\Bigr) \label{dilog} 
\end{eqnarray}
At small $x$ we have $L(x) = x - \frac{x^2}{4} + \frac{x^3}{9} - \frac{x^4}{16} + \ldots$. This power-type expansion and Eq.(\ref{dilog}) allows one to determine the asymptotics of the cross-section 
$\sigma_r$ at small and very large energies of the incident electron. In particular, at small energies, when $E_1 \ll m$, one finds $\sigma_r = \frac{16}{3} \alpha^5 a^2_0 \Bigl(\frac{Q^2}{\omega}\Bigr) 
\Bigl[ 1 - \frac{F(q)}{Q} \Bigr]^2$. In this case the radiative cross-section does not depend explicitly upon the energy of the incident electron $E_1$. At large energies of the incident electron, 
when $E_1 \gg m$, the analogous formula for multicharged ions is written in the form $\sigma_r = 4 \alpha^5 a^2_0 \Bigl(\frac{Q^2}{\omega}\Bigr) \Bigl[ 1 - \frac{F(q)}{Q} \Bigr]^2 \Bigl[\ln\Bigl(\frac{2 
E_1}{m}\Bigr) - \frac13\Bigr]$. For neutral atoms one needs to apply a different formula, since in this case the bremsstrahlung cross-section converges to the constant limit when $E_1 \rightarrow \infty$ 
and does not include any term which logarithmically diverges at large energies. 

Actual high-temperature plasma consists of multicharged ions and free electrons. Therefore, a number of other processes also contribute to the energy losses in actual experimental systems. The most 
important of such processes is the non-eleastic electron scattering by atoms and positively charged ions. Another process is electron-electron scattering which leads to energy loss of the fast electron 
and accelerations of many `secondary' electrons, or $\delta-$electrons. A third process is the Compton scattering of photons by free electrons with the emission of different photons. In the first order 
this later process is strictly prohibited in QED, but in the second order it is possible. All these processes which compete with the bremsstrahlung have been analyzed in the literature. The energy loss 
(per unit length of the trajectory) of the fast electron during its non-elastic collisions with atoms/ions can be evaluated from the following approximate formula \cite{Breit}:  
\begin{eqnarray}
     \Bigl(\frac{d E_1}{d x}\Bigr)_{i} = - 2 \pi \alpha^2 a^2_{0} N Q \ln\Bigl(\frac{\gamma^3_{1} m^2 c^4}{I}\Bigr)  \label{breit}
\end{eqnarray}
where $\gamma_1 = \frac{E_1}{m c^2}$ is the electron's $\gamma-$factor, $N$ is the number of ions/atoms per $cm^3$ and $I \approx$ 13.605 $eV$ is the `atomic' ionization potential. The ratio $R$ of the  
$-\Bigl(\frac{d E_1}{dx}\Bigr)_{r}$ and $-\Bigl(\frac{d E_1}{d x}\Bigr)_{i}$ derivatives is written in the form: $R = \frac{\gamma_1 Q}{A}$, where $A (\approx 1600)$ is a numerical constant which is 
chosen to produce the best fits for known experimental data.

Consider now the process of electron-electron scattering which leads to the formation of the secondary accelerated electrons. Let us designate by $\Delta$ the dimensionless ratio of the energy transfered 
by the fast electron to the secondary electron, i.e., $E_1 - E^{\prime}_1 (= E^{\prime}_2 - m)$, which was originally at rest, to the kinetic energy of the fast electron $E_1 - m c^2 = E_1 - m$. With this
notation we can write the following formula for the differential cross-section of the electron-electron scattering (see, e.g., \cite{AB})
\begin{eqnarray}
  \frac{d \sigma_{ee}}{d \Delta} = \frac{2 \pi \alpha^4 a^2_0}{v^2_{1} (\beta - 1)} \frac{1}{\Delta^2 (1 - \Delta)^2} \Bigl\{ 1 - \Bigl[3 - \Bigl(\frac{\beta - 1}{\beta}\Bigr)^2\Bigr] \Delta (1 - \Delta)
  + \Bigl(\frac{\beta - 1}{\beta}\Bigr)^2 \Delta^2 (1 - \Delta)^2\Bigr\} \label{sigmaee} 
\end{eqnarray}
where $\beta = \frac{E_1}{m}$ is the beta factor of the fast electron. It is interesting to note that the differential cross-section of the electron-electron scattering is a relatively simple function of 
the parameters $\beta$ and $\Delta (1 - \Delta)$. For small $\Delta$ one finds $\frac{d \sigma_{ee}}{d \Delta} = \frac{2 \pi \alpha^4 a^2_0}{v^2_{1} (\beta - 1) \Delta^2}$. In dense media the energy loss 
due to electron-electron scattering is in direct competition with the bremsstrahlung and non-elastic collisions of fast electron(s) with atoms/ions. 

The last process which we want to discuss here is the Compton scattering of photons by a free (fast) electron. Actual high-temperature plasmas contains a large number of photons which can interact with 
free electrons. Such a process leads to the emission of secondary photons. The energy of the secondary photon can be larger, or smaller than the energy of the original photon. In the first case we 
deal with the Compton scattering of photon by a fast electron. The differential cross-section of the Compton scattering is written in the form (see, e.g., \cite{Grein})
\begin{eqnarray}
  \frac{d \sigma_{C}}{d \omega_2} = \frac{2 \alpha^4 a^2_0 \omega^{2}_{2}}{m^2 \kappa^{2}_{1}} \Bigl[4 \Bigl(\frac{1}{\kappa_1} + \frac{1}{\kappa_2}\Bigr)^2 - 4 \Bigl(\frac{1}{\kappa_1} + 
  \frac{1}{\kappa_2}\Bigr) - \Bigl(\frac{\kappa_1}{\kappa_2} + \frac{\kappa_2}{\kappa_1}\Bigr)\Bigr] \label{sigcom}
\end{eqnarray}
where $\omega_2 = \hbar \omega_2$ is the energy of the secondary photon, while the parameters of the incident and final electrons $\kappa_1$ and $\kappa_2$ equal the products of the electron and photon 
4-vectors, i.e., $\kappa_1 = \frac{2 p_1 k_1}{m^2}$ and $\kappa_2 = - \frac{2 p_1 k_2}{m^2}$, respectively. As follows from the energy conservation law(s) for the Compton scattering, the same parameters 
$\kappa_1$ and $\kappa_2$ can be written in the forms: $\kappa_1 = \frac{2 p_2 k_2}{m^2}$ and $\kappa_2 = - \frac{2 p_2 k_1}{m^2}$. The frequency of the secondary photon can de evaluated from the 
following relation for the four-vectors: $p_1 + k_1 = p_2 + k_2$ (energy conservation law). Indeed, calculating the both sides of the equality $(p_1 + k_1)^2 = (p_2 + k_2)^2$ with the use of the 
conditions $p^2_1 = p^2_2 = -m^2$ and $k^2_1 = k^2_2 = 0$ one finds the following identity $p_1 k_1 = p_1 k_2 + k_1 k_2$, or in other words:
\begin{eqnarray}
   \omega_1 (1 - v_1 \cos\theta_1) = \omega_2 (1 - v_1 \cos\theta_2) + \frac{\omega_1 \omega_2}{E_1} (1 - v_1 \cos\theta)  \label{sigcom1}
\end{eqnarray}
where $v_1$ and $E_1$ are the velocity and energy of the incident electron, $\omega_1$ and $\omega_2$ are the frequencies of the incident and secondary photons, while $\theta_1$ and $\theta_2$ designate 
the angles between vectors ${\bf p}_1$ and ${\bf k}_1$ and vectors ${\bf p}_1$ and ${\bf k}_2$, respectively. Analogously, the angle $\theta$ is the angle between ${\bf k}_1$ and ${\bf k}_2$. More details
about such calculations can be found, e.g., in Section 3.7 of \cite{Grein}. Note that in high-temperature plasmas the Compton scattering of photons by fast electrons is one of the leading channels of energy 
loss. This process is always competing with the high-temperature bremsstrahlung.  
  
\subsection{Bremsstrahlung from high-temperature plasma}

Bremsstrahlung from hot plasma with temperatures $T \approx$ 7 - 15 $keV$ is of great interest for actual applications mainly related to nuclear fusion. In general, such a plasma can be considered
as a system at local thermal equilibrium. Therefore, we can investigate this by using a set of additional equations which follow from conditions of thermal equilibrium and allow us to obtain a number 
of relations between different properties of radiating plasma. In this case we can introduce separate temperatures for each plasma component, i.e. for ions $T_i$, electrons $T_e$ and radiation $T_r$. To 
simplify the description of bremsstrahlung and/or Compton scattering below we consider a two component plasma which contains only electrons and radiation with temperatures $T_e$ and $T_r$. It is clear 
that a hot plasma of light elements $Q \le 3$ with temperatures $T \approx$ 7 - 15 $keV$ contains only electrons and bare atomic nuclei. For such a plasma only bremsstrahlung and/or Compton scattering of 
electrons are important channels of energy loss. Without an extensive discussion of energy balance in fusion-related plasmas we just present the explicit formula for the coefficient $A_{er}$ which 
determines the energy transfer rate (or temperature transfer rate) betweeen electrons and radiation in a high-temperature plasma which is assumed to be at local thermal equilibrium. If $T_e$ and $T_r$ 
are the radiation and electron temperatures of the plasma, then we can write $A_{er} = C_{\nu e} (\nu_b + \nu_C)$, where $C_{\nu e} = \frac{16 \sigma}{c \rho} \cdot T^{3}_{r}$ \cite{Fra}, and the overall 
bremsstrahlung rate $\nu_b$ is
\begin{eqnarray}
 \nu_b = 8.510768649 \cdot \frac{k e^4 N^2_A}{\hbar c} \Bigl(\frac{Z^2}{A^2}\Bigr) \frac{\rho Z}{(m_e k T_e)^{\frac12}} G\Bigl(\frac{T_r}{T_e}\Bigr) \label{bremrate}
\end{eqnarray}
where $N_A = 6.02214179 \cdot 10^{23}$ is the Avogadro number, $\sigma$ is the Stephan-Boltzmann constant, $c$ is the speed of light in vacuum, $m_e$ and $e$ are the electron mass and electric charge 
of the electron, respectively, while $k$ is the Boltzmann constant and $G(x)$ is the following function
\begin{eqnarray}
   G(x) = \frac{1}{x - 1} \int_{0}^{+\infty} \frac{d\xi f(\xi) \{1 - \exp[-\xi (\frac{1}{x} - 1)]\}}{1 - \exp(-\frac{\xi}{x})}
\end{eqnarray} 
where $f(\xi)$ is the function
\begin{eqnarray}
   f(\xi) = \exp(\xi) \int_{0}^{+\infty} \ln\Bigl(\sqrt{y + 1} + \sqrt{y}\Bigr) \exp(-\xi y) dy
\end{eqnarray}
The formula, Eq.(\ref{bremrate}), determines the `pure' bremsstrahlung rate for $T_r < T_e$ and inverse bremsstrahlung rate, if $T_e < T_r$. The Compton scattering rate $\nu_C$ is \cite{Fra}
\begin{eqnarray}
 \nu_C = 134.0412866 \cdot \frac{e^2 \sigma}{(m_e c^2)^2} N_A \Bigl(\frac{Z}{A}\Bigr) k T^4_r \label{Comptrate}
\end{eqnarray}
Note that the product of the Boltzmann constant and Avogadro number equals the universal gas constant $R$, i.e.  $k N_A = R$.

These formulas can be used to determine a heat propagation in the light-element plasma from an initial (hot) area into new (cold) areas of the plasma. In general, if energy losses during such a heat 
propagation into cold areas are over-compensated by the energy release from thermonuclear reactions, then this plasma ignites, or burns-up. In actual application the high-temperature burn-up of 
deuterium plasmas which contain significant amounts of deuterium and tritium nuclei and which can also be mixed with some amounts of the ${}^{4}$He and ${}^{6}$Li nuclei, is of great interest.   
Moreover, it is clear that the burn-up in high-temperature plasma will be optimal if the nuclear fusion reactions start in one very small spatial area (or `point') and propagate from this hot point 
to other areas occupied by the cold thermonuclear fuel.    

\subsection{Bremsstrahlung role in the high-temperature burn-up of the deuterium plasma}

Let us assume that inside of an infinite thermonuclear fuel with density $\rho_0$, part of the fuel in volume ${\cal V}$ is instantaneously heated to a high temperature $T$. If the values of ${\cal V}$ and 
$T$ exceed some critical values ${\cal V}_c$ and $T_c$ then thermonuclear burning begins in the volume ${\cal V}_c$ and such a burning wave can propagate to the rest of the fuel \cite{Avr} - \cite{Fro00}. 
This corresponds to the burn-up of thermonuclear fuel from a central hot point. In spherical geometry this process is governed by the following equation \cite{Avr}
\begin{eqnarray}
   C \frac{d T}{d t} = - C \frac{3}{r_f} \frac{d r_f}{d t} T + q(r_f, \rho, T) \; \; \; or  \; \; \;  \frac{d T}{d x} = - \frac{3}{x} T + \frac{Q(x, \rho, T)}{c V_{max}} \label{burnup}
\end{eqnarray}
where $C$ is the specific heat per unit mass of the fuel, $r_f(t)$ is the radius of the hot zone (also called the combustion zone), $x = \rho_0 r_f$ is the burn-up parameter and $d x = \rho_0 dr_f = 
\rho_0 V_{max} dt$, where $V_{max}$ is the speed of burn wave propagation (at givien $\rho_0$ and $T$), or velocity of the hot zone expansion. The notation $Q(x, \rho, T)$ in Eq.(\ref{burnup}) stands for 
the energy release function which depends upon chemical composition of the fuel, burn-up parameter $x$ and density $\rho$. In actual high-temperature plasmas the burning wave can propagate in a few 
different ways, e.g., by compressing initially cold fuel by a very strong shock wave (or by a consequence of such waves), or by high-temperature heat conduction. The first way corresponds to the 
high-temperature detonation \cite{Feok}, while the second way represents a high-temperature electron heat wave. In some plasmas at certain conditions a few other ways of heat propagation are also possible, 
e.g., by fast $\alpha-$particles (or other electrically charged fast particles) formed in nuclear fusion reactions and/or fast neutrons which are often formed in such reactions. Below, we consider only 
actual plasmas in which propagation of the hot zone proceeeds by high-temperature detonation and/or high-temperature thermal conductivity.

The crucial question about any possible thermonuclear fuel is the explicit form of the burn-up function $Q(x, \rho, T)$ in Eq.(\ref{burnup}). For instance, for equimolar deuterium-tritium mixture this 
function takes the form \cite{Avr}
\begin{eqnarray}
  Q(x, \rho, T) &=& (1 + 4 k_n) \cdot 2.17 \cdot 10^{7} \cdot \frac{1 + 0.232 T^{0.75}}{T^{\frac23} \sqrt{1 + 9.4 \cdot 10^{-5} T^{3.25}}} \cdot \exp(-\frac{20}{T^{\frac23}}) \nonumber \\
  &-& \frac{31 \sqrt{T}}{1 + 1.1 \sqrt{x \rho} T^{-1.75}} \label{Qfunct} 
\end{eqnarray}
where $k_n = \frac{0.048 x}{1 + 0.048 x}$ is that protion of the energy which neutrons are leaving in a sphecial ball of radius $r$ located inside of the equimolar DT-plasma with density $\rho$. The first term 
in Eq.(\ref{Qfunct}) is the product of the fusion reaction cross-section and velocity of the colliding nuclei ($\sigma v$) averaged over Boltzmann energy distribution. For the $(d,t)-$nuclear reaction this
(first) term is very large, since the energy realease from this reaction is huge and it has a very large resonance at $T \approx 107$ $keV$. For a pure deuterium plasma, the analogous term is substantially 
smaller (59 - 78 times smaller depending on temperature). The explicit formula of the first term in  Eq.(\ref{Qfunct}) for the $(d,d)-$reaction can be found, e.g., in \cite{Glasst}. Therefore, for pure 
deuterium plasmas the second term in  Eq.(\ref{Qfunct}) becomes important. Breifly, the second term represents energy loss from the reaction (hot) zone due to flux of high-temperature radiation. This term 
is often called the bremsstrahlung (negative) contribution. In earlier astrophysical studies performed 70 - 75 years ago for hydrogen plasmas this term was written in the form $- 31\sqrt{T}$. However, such 
a form for the bremsstrahlung term in Eq.(\ref{Qfunct}) leads to a uniform conclusion that ignition of the deuterium plasmas in finite volumes with spatial radius less than 1 meter is not possible. Formally, 
thermonuclear ignition is possible when spatial radius of the ball of deuterium plasma exceeds 800 meters, but it is absolutely non-realistic in applications. After extensive reasearch of deuterium plasmas it 
became clear that all radiation quanta emitted from the hot zone due to high-temperature bremsstrahlung cannot reach the bondaries of the hot zone. This means that some part of these quanta is absorbed in 
this hot zone. In general, the number of absorbed quanta is directly proportional to the density $\rho$ of the plasma and burn-up parameter $x$. The factor $\Bigl( 1 + 1.1 \cdot \sqrt{x \rho} \cdot T^{-1.75} 
\Bigr)^{-1}$ in the second term of Eq.(\ref{Qfunct}) represents the bremsstrahlung screening. Note that this factor contains two control parameters (density and linear size). Variation of these parameters and 
first of all density $\rho$ allows one to produce ignition of pure deuterium and other thermonuclear plasmas. This explains a crucial role of high-temperature bremsstrahlung for ignition of various 
deuterium-containing plasmas.               

\section{Conclusion} 

We developed an approach which can be used to evaluate contributions of the electron-electron correlations into bremsstrahlung cross-section and corresponding energy loss for few-electron ions and atoms. The 
crucial part of accurate atomic calculatons is to determine the form-factors $F(q)$ of the incident few-electron ions/atoms, i.e. the Fourier-transform of the one-electron density distribution. In the central 
field approximation (which is very good for these problems) actual computations are reduced to the accurate computation of a few one-dimensional integrals. Such computations have been performed for all 
two-electron ions/atoms (He - Ni$^{26+}$) yielding the closed analytical expression for all matrix elements which are needed to determine atomic form-factors of all two-electron atoms and ions to very high 
accuracy within a central field approximation in which the electron distribution density is considered to be a spherically symmetric. There is a possibility to generalize our approach to three- and four-electron 
atoms/ions and even to the case of atoms/ions with arbitrary number of bound electrons. We also discuss the energy loss due to bremsstrahlung from high-temperature plasmas which contain multi-charged ions and 
free electrons. It is shown that screening of bremsstrahlung in high-temperature plasma plays a central role in the burn-up of high-temperature deuterium plasmas.

\newpage
\begin{table}[tbp]
   \caption{Convergence of the atomic form-factor in $a.u.$ of the He-atom (the ground $1^1S-$states) upon the total number of basis functions used. Atomic nucleus is 
            assumed to be infinitely heavy.}
     \begin{center}
     \scalebox{0.90}{%
     \begin{tabular}{| c | c | c | c | c | c |}
      \hline\hline
   $q$  & $F(q)$ ($N$ = 2000) &  $F(q)$ ($N$ = 2500) & $F(q)$ ($N$ = 3000) & $F(q)$ ($N$ = 3500) & $F(q)$ ($N$ = 4000) \\ 
          \hline    
   0.1 & 1.99602833554455 & 1.99602833554456 & 1.99602833554457 & 1.99602833554456 & 1.99602833554456 \\
   0.2 & 1.96472444168180 & 1.96472444168181 & 1.96472444168181 & 1.96472444168180 & 1.96472444168180 \\
   0.3 & 1.93800126413211 & 1.93800126413213 & 1.93800126413214 & 1.93800126413212 & 1.93800126413213 \\
   0.4 & 1.90452526569111 & 1.90452526569110 & 1.90452526569111 & 1.90452526569110 & 1.90452526569110 \\ 
   2.5 & 0.80608623318909 & 0.80608623318908 & 0.80608623318909 & 0.80608623318908 & 0.80608623318908 \\
   2.7 & 0.72156548850375 & 0.72156548850375 & 0.72156548850375 & 0.72156548850376 & 0.72156548850375 \\
   2.8 & 0.68250273317700 & 0.68250273317701 & 0.68250273317702 & 0.68250273317701 & 0.68250273317701 \\
   5.7 & 0.01486064557919 & 0.01486064557918 & 0.01486064557919 & 0.01486064557918 & 0.01486064557918 \\
   5.8 & 0.01417307437410 & 0.01417307437411 & 0.01417307437411 & 0.01417307437412 & 0.01417307437411 \\
   6.0 & 0.01290641070551 & 0.01290641070550 & 0.01290641070551 & 0.01290641070550 & 0.01290641070550 \\
    \hline\hline
  \end{tabular}}
  \end{center}
  \end{table}
\newpage
\begin{table}[tbp]
   \caption{Atomic form-factors in $a.u.$ for some two-electron ions (ground $1^1S-$states). Atomic nuclei are assumed to be infinitely heavy.}
     \begin{center}
     \scalebox{0.75}{%
     \begin{tabular}{| c | c | c | c | c | c |}
      \hline\hline
   $q$  & $F(q)$ (H$^{-}$) &  $F(q)$ (He) & $F(q)$ (Ne$^{8+}$) & $F(q)$ (Ca$^{18+}$) & $F(q)$ (Ni$^{26+}$) \\ 
          \hline   
   0.0 & 2.0                       & 2.0              &  2.0             &   2.0            & 2.0    \\   
   0.1 & 1.96132945057312 & 1.99602833554456    & 1.999891483947858 & 1.999973970809488 & 1.999986873353467 \\
   0.2 & 1.85640934479775 & 1.96472444168180    & 1.999565989597304 & 1.999895886309433 & 1.999947494193332 \\
   0.3 & 1.71043482420304 & 1.93800126413213    & 1.999023678306149 & 1.999765755713475 & 1.999881864857887 \\
   0.4 & 1.54833795764938 & 1.90452526569110    & 1.998264818804848 & 1.999583594374983 & 1.999789989243940 \\ 
   0.5 & 1.38717714574940 & 1.86490379021588    & 1.997289786898069 & 1.999349423783013 & 1.999671872806301 \\
   0.6 & 1.23602867177643 & 1.81982405765591    & 1.996099065047703 & 1.999349423783013 & 1.999527522557058 \\  
   0.7 & 1.09860493965672 & 1.77002688319127    & 1.994693241838125 & 1.999063271556655 & 1.999356947064652 \\
   0.8 & 0.97567464681582 & 1.71628065247207    & 1.993073011324763 & 1.998335163282062 & 1.999160156452743 \\
   1.0 & 0.77020482817221 & 1.65935700342965    & 1.989192627248489 & 1.997399612788159 & 1.998687978132901 \\ 
           \hline
   2.0 & 0.25140513353668  & 1.05558832506606   & 1.957298839612961 & 1.989629032969506 & 1.994759690186353 \\  
   2.1 & 0.22631640202285  & 1.00140920768267   & 1.953000585093706 & 1.988570603016670 & 1.994223728568467 \\
   2.2 & 0.20400578198386  & 0.94931116662983   & 1.948508086746243 & 1.987461456215148 & 1.993661855738936 \\
   2.3 & 0.18414561608885  & 0.89936918205287   & 1.943823468832696 & 1.986301721988844 & 1.993074104874982 \\ 
   2.4 & 0.16644818561710  & 0.85162372319241   & 1.938948936453435 & 1.985091535520344 & 1.992460510663290 \\
   2.5 & 0.15066077068448  & 0.80608623318908   & 1.933886773267480 & 1.983831037715909 & 1.991821109295450 \\
   2.6 & 0.13656133733834  & 0.76274416108052   & 1.928639339147500 & 1.982520375169048 & 1.991155938463219 \\
   2.7 & 0.12395476256023  & 0.72156548850375   & 1.923209067773552 & 1.981159700122672 & 1.990465037353574 \\
   2.8 & 0.11266952478171  & 0.68250273317701   & 1.917598464169758 & 1.979749170429865 & 1.989748446643585 \\  
   2.9 & 0.10255480002725  & 0.64549643651892   & 1.911810102188170 & 1.978288949513273 & 1.989006208495098 \\
   3.0 & 0.093477913249790 & 0.61047816022939   & 1.905846621944145 & 1.976779206323141 & 1.988238366549214 \\  
          \hline
   5.0 & 0.0019119691603341 & 0.02092499944253  & 1.754339984527224 & 1.936491784524877 & 1.967584769495008 \\ 
   5.1 & 0.0017868988736314 & 0.01990385784924  & 1.745397409937382 & 1.933990377128254 & 1.966291880377667 \\ 
   5.2 & 0.0016716011208328 & 0.01893974669421  & 1.736349044142877 & 1.931444462097725 & 1.964974708365103 \\ 
   5.3 & 0.0015652000117109 & 0.01802919975723  & 1.727198296048866 & 1.928854321129029 & 1.963633329186683 \\ 
   5.4 & 0.0014669071827418 & 0.01716896720854  & 1.717948575709361 & 1.926220240087431 & 1.962267819865867 \\ 
   5.5 & 0.0013760122531516 & 0.01635600319142  & 1.708603291878175 & 1.923542508939155 & 1.960878258710472 \\ 
   5.6 & 0.0012918743968055 & 0.01558745381000  & 1.699165849608033 & 1.920821421682095 & 1.959464725302800 \\  
   5.7 & 0.0012139148921932 & 0.01486064557918  & 1.689639647900089 & 1.918057276275817 & 1.958027300489612 \\  
   5.8 & 0.0011416105304558 & 0.01417307437411  & 1.680028077405945 & 1.915250374570904 & 1.956566066371970 \\  
   5.9 & 0.0010744877767607 & 0.01352239490186  & 1.670334518184135 & 1.912401022237641 & 1.955081106294943 \\  
   6.0 & 0.0010121175936764 & 0.01290641070550  & 1.660562337512848 & 1.909509528694104 & 1.953572504837169 \\  
         \hline
   10.0 & 0.0001516730439791 & 0.002624513248630 & 1.24042038231789 & 1.763227052540513 & 1.874953401633500 \\  
   10.1 & 0.0001459992755751 & 0.002536906728319 & 1.22994632121624 & 1.758910183840519 & 1.872561989249619 \\  
   10.2 & 0.0001405861697800 & 0.002452820055244 & 1.21950396802771 & 1.754566548761648 & 1.870151446564242 \\ 
   10.3 & 0.0001354196056549 & 0.002372089736317 & 1.20909496362771 & 1.750196545744651 & 1.867721899795742 \\
   10.4 & 0.0001304863393311 & 0.002294560815708 & 1.19872090064025 & 1.745800573630203 & 1.865273475839531 \\
   10.5 & 0.0001257739428235 & 0.002220086384021 & 1.18838332376428 & 1.741379031587553 & 1.862806302254074 \\
   10.6 & 0.0001212707475630 & 0.002148527118019 & 1.17808373013037 & 1.736932319043730 & 1.860320507246883 \\
   10.7 & 0.0001169657922505 & 0.002079750848878 & 1.16782356968615 & 1.732460835613317 & 1.857816219660491 \\ 
   10.8 & 0.0001128487746731 & 0.002013632157083 & 1.15760424560905 & 1.727964981028818 & 1.855293568958419 \\ 
   10.9 & 0.0001089100071503 & 0.001950051992192 & 1.14742711474504 & 1.723445155071623 & 1.852752685211118 \\
    \hline\hline
  \end{tabular}}
  \end{center}
  \end{table}
%


\begin{table}[tbp]
   \caption{The electron-nuclear $\langle r^{2 k}_{eN} \rangle$ expectation values (in $a.u.$) for a number of two-electron ions
            with infinitely heavy atomic nuclei.} 
     \begin{center}
     \scalebox{0.72}{%
     \begin{tabular}{| c | c | c | c | c | c |}
      \hline\hline
  ion/atom &      H$^{-}$       &        He         &       Ne$^{8+}$             & Ca$^{18+}$ &         Ni$^{26+}$        \\ 
     \hline
 $\langle r^{2}_{eN} \rangle$  & 11.913699678051262 & 1.19348299501893527 & 3.2556160988739701$\cdot 10^{-2}$ & 7.8088339435014055$\cdot 10^{-3}$ & 3.93801344691995127$\cdot 10^{-3}$ \\
 
 $\langle r^{4}_{eN} \rangle$  & 645.1445424122194  & 3.9735649316629101  & 2.690792104845699$\cdot 10^{-3}$  & 1.5358087156342157$\cdot 10^{-4}$ & 3.8974100557929396$\cdot 10^{-5}$ \\
 
 $\langle r^{6}_{eN} \rangle$  & 87266.1424069593   & 26.28244697552571   & 4.188297920816449$\cdot 10^{-4}$  & 5.6625739359113182$\cdot 10^{-6}$ & 7.222000786090979$\cdot 10^{-7}$ \\

 $\langle r^{8}_{eN} \rangle$  & 22035718.695491    & 289.827674593716    & 1.05338028889520$\cdot 10^{-4}$   & 3.364163448868867$\cdot 10^{-7}$  & 2.15474156619540$\cdot 10^{-8}$ \\
  
 $\langle r^{10}_{eN} \rangle$ & 8937970758.421     & 4797.7265440809     & 3.8992426943743$\cdot 10^{-5}$    & 2.93640072611717$\cdot 10^{-8}$   & 9.5023197194145$\cdot 10^{-10}$ \\
     \hline \hline
  \end{tabular}}
  \end{center}
  \end{table}

\newpage  

\begin{thebibliography}{10}

\bibitem{Heitl} W. Heitler, {\it The Quantum Theory of Radiation}, (3rd ed., Oxford at the Clarendon Press, London, 1954), \S 25. 

\bibitem{CRC} \textit{CRC Handbook of Chemistry and Physics}, 92nd Edition, Ed. W.M. Haynes, (Taylor and Francis Group, Boca Raton, FL, 2011-2012).

\bibitem{Grein} W. Greiner and J. Reinhardt, {\it Quantum Electrodynamics}, (4th edn., Springer Verlag, Berlin, 2009).

\bibitem{AB} A.I. Akhiezer and V.B. Beresteskii, {\it Quantum Electrodynamics}, (4th ed., Science, Moscow (1981)), Chpt. 4 and 5.

\bibitem{LLQ} L.D. Landau and E.M. Lifshitz, {\it Quantum Mechanics. Non-Relativistic Theory}, (3rd. ed., Oxford, England, Pergamon Press, 1977).

\bibitem{Bethe} H.A. Bethe, {\it Intermediate Quantum Mechanics}, (New York, W.A. Benjamin, 1964).

\bibitem{GR} I.S. Gradstein and I.M. Ryzhik, \textit{Tables of Integrals, Series and Products}, (6th revised ed., Academic Press, New York, 2000).

\bibitem{Fro98} A.M. Frolov, Phys. Rev. A {\bf 57} (1998) 2436.

\bibitem{Fro2001} A.M. Frolov, Phys. Rev. E {\bf 64} (2001) 036704; ibid, {\bf 74} (2006) 027702.

\bibitem{Fro2015} A.M. Frolov, Chem. Phys. Lett. {\bf 638}, 312 - 320 (2015).

\bibitem{MotMessi} N.F. Mott and H.S.W. Messi, \textit{The Theory of Atomic Collisions}, (3rd revised ed., Oxford at the Clarendon Press, London, 1965).

\bibitem{Abram} Handbook of Mathematical Functions, Eds. M. Abramowitz and I.A. Stegun (Dover, New York, 1972).

\bibitem{Wolfr} http://mathworld.wolfram.com/Dilogarithm.html

\bibitem{Breit} G. Breit and E. Teller, Astrophys. Journal {\bf 91}, 215 (1940).


\bibitem{Fra} G.S. Fraley, E.J. Linnebur, R.J. Mason and R.L. Morse, Phys. Fluids {\bf 17}, 474 (1974).

\bibitem{Avr} E.N. Avrorin, L.P. Feoktistov, and L.I. Schibarshov. Fiz. Plasmy {\bf 6}, 965 (1980) [Engl. transl. Sov. J. Plasma Phys. {\bf 6}, 527 (1980)].

\bibitem{Feok} L.P. Feoktistov, Uspekh. Fiz. \textit{Thermonuclear detonation}, {\bf 168}, 1247 (1998) [Engl. transl. Sov. Phys. Uspekhi. {\bf 168}, 1100 (1998)].

\bibitem{Fro00} A.M. Frolov, Phys. Rev. E {\bf 62}, 4104 (2000).

\bibitem{Glasst} S. Glasstone and R.H. Lovberg. \textit{Controlled thermonuclear reactions} (Van Nostrand, Princeton, N.J. 1960).

%
\end{thebibliography}
\end{document}